\documentclass[pra,twocolumn,showpacs]{revtex4-1}
\usepackage{amsmath,amscd,amsfonts,amssymb,color}
\usepackage{graphicx,amsfonts,epsf}
\usepackage[pdftex]{hyperref}
\begin{document}
\title{Experimental protection 
of arbitrary states in a two-qubit
subspace by nested Uhrig dynamical decoupling}
\author{Harpreet Singh}
\email{harpreetsingh@iisermohali.ac.in}
\affiliation{Department of Physical Sciences, Indian
Institute of Science Education and 
Research Mohali, Sector 81 SAS Nagar, 
Punjab 140306 India.}
\author{Arvind}
\email{arvind@iisermohali.ac.in}
\affiliation{Department of Physical Sciences, Indian
Institute of Science Education  and
Research Mohali, Sector 81 SAS Nagar, 
Punjab 140306 India.}
\author{Kavita Dorai}
\email{kavita@iisermohali.ac.in}
\affiliation{Department of Physical Sciences, Indian
Institute of Science Education and 
Research Mohali, Sector 81 SAS Nagar, 
Punjab 140306 India.}
\begin{abstract}
We experimentally demonstrate the efficacy of a three-layer
nested Uhrig dynamical decoupling (NUDD) sequence to
preserve arbitrary quantum states in a two-dimensional
subspace of the four-dimensional two-qubit Hilbert space, on
an NMR quantum information processor. The effect of the
state preservation is studied first on four known states,
including two product states and two maximally entangled
Bell states.  Next, to evaluate the preservation capacity of
the NUDD scheme, we apply it to eight randomly generated
states in the subspace. Although, the preservation of
different states varies, the scheme on the average performs
very well.  The complete tomographs of the states at
different time points are used to compute fidelity.  State
fidelities using NUDD protection are compared with those
obtained without using any protection.  The nested pulse
schemes are complex in nature and require careful
experimental implementation.
\end{abstract}
\maketitle
\section{Introduction}
\label{intro}
Dynamical decoupling (DD) sequences have found widespread
application in quantum information processing (QIP), as
strategies for protecting quantum states against
decoherence~\cite{viola-review}.  For a quantum system
coupled to a bath, the DD sequence decouples the system and
bath by adding a suitable decoupling interaction, periodic
with cycle time $T_c$ to the overall system-bath 
Hamiltonian~\cite{viola-prl-99-2}.  After $N$ applications of
the cycle for a time $N T_c$, the system is governed by a
stroboscopic evolution under an effective average
Hamiltonian, in which system-bath interaction
terms are no longer present.

The simplest DD sequences were motivated by early NMR
spin-echo based schemes for coherent averaging of unwanted
interactions~\cite{carr-pr-54}, and used  periodic
time-symmetrized trains of instantaneous $\pi$ pulses
(equally spaced in time) to suppress decoherence.  More
sophisticated DD schemes are of the Uhrig dynamical
decoupling (UDD) type, wherein the pulse timing in the DD
sequence is tailored to produce higher-order cancellations
in the Magnus expansion of the effective average
Hamiltonian,
thereby achieving system-bath decoupling to a
higher order and hence stronger noise
protection~\cite{uhrig-njp-08,hodgson-pra-10,schroeder-pra-11,yang-fpc-11,liu-nc-13}.
UDD schemes are applicable when the control pulses can be
considered as ideal (i.e. instantaneous) and when the
environment noise has a sharp frequency
cutoff~\cite{dhar-prl-06,yang-prl-08,uhrig-prl-09,khodjasteh-pra-11}.
These initial UDD schemes dealt with protecting a single
qubit against different types of noise, and were later
expanded to a whole host of optimized sequences involving
nonlocal control operators, to protect multi-qubit systems
against
decoherence~\cite{mukhtar-pra-10-1,pan-jpb-11,cong-ijqi-11,alvarez-pra-12,west-njp-12,ahmed-pra-13}.
Quantum entanglement is considered to be a crucial resource
for QIP, and several studies have explored the efficacy of
UDD protocols in protecting such fragile quantum
correlations against
decay~\cite{agarwal-scripta,song-ijqi-13,lofranco-prb-2014}.
The experimental performance of UDD schemes have been
demonstrated for trapped ion qubits undergoing
dephasing~\cite{biercuk-pra-09,szwer-jpb-11}, for electron
spin qubits decohering in a spin bath~\cite{du-nature-09}
and for  NMR
qubits~\cite{alvarez-pra-10,ajoy-pra-11,roy-pra-2011}.  The
freezing of state evolution using super-Zeno sequences was
experimentally demonstrated using NMR~\cite{singh-pra-14},
and DD sequences were interleaved with quantum gate
operations in an electron-spin qubit of a single
nitrogen-vacancy center in diamond~\cite{zhang-prl-14}.
Non-QIP applications of DD schemes include their usage for
enhanced contrast in magnetic resonance imaging of tissue
samples~\cite{jenista-jcp-09} and for suppression of NMR
relaxation processes whilst studying molecular diffusion via
pulsed field gradient experiments~\cite{alvarez-jcp-14}.

While UDD schemes can well protect states against 
single- and two-axis noise (i.e. pure dephasing and/or pure bit-flip),
they are not able to protect against general three-axis
decoherence~\cite{kuo-jmp-12}.
Nested UDD (NUDD) schemes were hence proposed to
protect multiqubit systems in generic quantum baths to
arbitrary decoupling orders,
by nesting several UDD layers and 
it was  shown that the NUDD scheme can preserve a set of
unitary Hermitian system operators (and hence all operators
in the Lie algebra generated from this set of operators)
that mutually either commute or anticommute~\cite{wang-pra-11}.
Furthermore,
it was proved that the NUDD scheme is universal i.e. it can
preserve the coherence of $m$ coupled qubits by suppressing
decoherence upto order $N$, independent of the nature of the
system-environment coupling~\cite{jiang-pra-11}.
Recently, a theoretical proposal examined in detail the
efficiency of NUDD schemes in 
protecting unknown randomly generated two-qubit 
states and showed that such schemes are a powerful approach
for protecting quantum states against
decoherence~\cite{mukhtar-pra-10-2}.

This work focuses on the preservation of arbitrary states in
a known two-dimensional subspace using appropriate NUDD
sequences on an NMR quantum information processor.  We first
evaluate the efficacy of protection of the NUDD scheme by
applying it on four specific states of the subspace ${\cal
P}=\{\vert 01 \rangle, \vert 10 \rangle \}$ i.e.  two
separable states: $|01\rangle$ and $|10\rangle$, and two
maximally entangled singlet and triplet Bell states:
$\frac{1}{\sqrt{2}}(|01\rangle-|10\rangle)$ and
$\frac{1}{\sqrt{2}}(|01\rangle+|01\rangle)$ in a
four-dimensional two-qubit Hilbert space.  Next, to evaluate
the effectiveness of the NUDD scheme on the entire subspace,
we randomly generate states in the subspace ${\cal P}$
(considered as a superposition of the known basis states
$\vert 01 \rangle, \vert 10 \rangle$) and protect them using
NUDD.  We randomly generate eight states in the two-qubit
subspace and protect them using a three-layer NUDD sequence.
Full state tomography is used to compute the experimental
density matrices.  We allow each state to decohere, and
compute the state fidelity at each time point without
protection and after NUDD protection.  The results are
presented as a histogram and show that while NUDD is always
able to provide some protection, the degree of protection
varies from state to state.

This is the first experimental demonstration of the efficacy
of NUDD sequences in protecting arbitrary states in a
two-qubit subspace against arbitrary noise, upto a
high-order.  Although NUDD schemes are designed to be
independent of any noise-model assumptions and also do not
require {\em a priori} information about the state to be
protected, they are experimentally challenging to implement
as they involve repeating cycles of several dozen rf pulses.
Nevertheless, their efficacy in suppressing decoherence to
higher orders in multiqubit systems makes them promising
candidates for realistic QIP.  Our experiments are hence an
important step forward in the protection of general quantum
states against general decoherence.

This paper is organized as follows:~Section~\ref{theory}
briefly recapitulates the NUDD scheme for two qubits and
gives details of how the nesting of three layers of UDD is
constructed in order to protect the diagonal populations and
the off-diagonal coherences against decoherence. The
explicit quantum circuit and corresponding NMR pulse
sequences to implement NUDD on two qubits is given in
Section~\ref{expt1}. The results of experimentally
protecting four specific states in the known subspace are
described in Section~\ref{known}.  Section~\ref{unknown}
contains a detailed description of the NUDD protection of a
randomly generated set of arbitrary states in the subspace
of two NMR qubits.  Finally, Section~\ref{concl} offers some
concluding remarks.
\section{The NUDD scheme}
\label{theory}
Consider a two-qubit quantum system with its state
space spanned by the states $\{ \vert 00 \rangle, \vert 01
\rangle, \vert 10 \rangle, \vert 11 \rangle \}$, the
eigenstates of the Pauli operator $\sigma_{z}^{1} \otimes 
\sigma_{z}^{2}$.
Our interest is in protecting states in the subspace
$\cal{P}$ spanned by states $\{\vert 01 \rangle,\vert 10
\rangle\}$, against decoherence.
The density matrix 
corresponding to an arbitrary pure state $\vert \psi \rangle
=\alpha \vert 01 \rangle + \beta \vert 10 \rangle$
belonging to the subspace $\cal{P}$
is given by
\begin{equation}
\rho(t)= \left(\begin{array}{cccc}
0&0&0&0 \\
0& \vert \alpha \vert^2&\alpha \beta^{*}&0 \\
0& \beta \alpha^{*}& \vert \beta \vert^2&0 \\
0&0&0&0
\end{array}
\right)
\end{equation}
with the coefficients $\alpha$ and $\beta$ satisfying 
$ \vert \alpha \vert^2+ \vert \beta \vert^2=1$ at 
time $t=0$.
We briefly describe here the theoretical construction
of a three-layer NUDD scheme to protect arbitrary states
in the two-qubit subspace $\cal{P}$~\cite{mukhtar-pra-10-1,mukhtar-pra-10-2}.

The general total Hamiltonian of a two-qubit system 
interacting with an arbitrary bath can be written as
\begin{equation}
H_{\rm total} =H_{S}+ H_{B} + H_{jB} + H_{12}  
\label{totham1}
\end{equation}
where $H_S$ is the system Hamiltonian, $H_{B}$ is
the bath Hamiltonian, $H_{jB}$ is qubit-bath
interaction Hamiltonian and $H_{12}$ is the
qubit-qubit interaction Hamiltonian (which can be
bath-dependent). 
Our interest here is in bath-dependent terms and
their control, which can be expressed using a
special basis set for the two-qubit system as
follows~\cite{mukhtar-pra-10-1,mukhtar-pra-10-2}:
\begin{eqnarray}
H &=& H_{B} + H_{jB} + H_{12}  \nonumber \\
  &=& H_0 + H_1 \nonumber \\
H_0 &=& \sum_{j=1}^{10} W_j Y_j, \,\, 
H_1 = \sum_{j=11}^{16} W_j Y_j 
\label{newbasis}
\end{eqnarray}
where the coefficients $W_j$ contain arbitrary bath
operators.  
$Y$ are the special basis computed from the
perspective of preserving the subspace spanned
by the states $\{ \vert 01 \rangle, \vert 10 \rangle \}$
in the two-qubit
space~\cite{mukhtar-pra-10-1,mukhtar-pra-10-2}:
\begin{eqnarray}
&&Y_{1}=I,
\quad 
\quad 
\quad 
\quad 
\,\,\,\,\,
Y_{2}=|01\rangle\langle01|+|10\rangle\langle10|, 
\nonumber \\
&&Y_{3}=
|00\rangle\langle11|,
\quad 
\quad
Y_{4}=
|00\rangle\langle00|-|11\rangle\langle11|,
\nonumber \\
&&Y_{5}=|11\rangle\langle00|,
\quad
\quad
Y_{6}=|01\rangle\langle01|-|10\rangle\langle10|, 
\nonumber \\
&&Y_{7}=|10\rangle\langle00|, 
\quad
\quad
Y_{8}=|00\rangle\langle10|,
\nonumber \\
&&Y_{9}=|10\rangle\langle11|,
\quad
\quad
Y_{10}=|11\rangle\langle10|,
\nonumber \\
&&Y_{11}=|01\rangle\langle00|,
\quad \,\,\,\,
Y_{12}=|00\rangle\langle01|,
\nonumber \\
&&Y_{13}=|01\rangle\langle11|,
\quad \,\,\,\,
Y_{14}=|11\rangle\langle01|, 
\nonumber \\
&&Y_{15}=|01\rangle\langle10|+|10\rangle\langle
01|,  \nonumber \\
&&Y_{16} =-i(|10\rangle\langle01|-|01\rangle\langle10|).
\end{eqnarray}

The recipe to design UDD protection for a two-qubit state
(say $\vert \chi \rangle$) is given in the following
steps:~(i) First a control operator $X_c$ is constructed
using $X_c = I-2 \vert \chi \rangle\langle \chi \vert$ such
that $X_c^2=I$, with the commuting relation $[X_c,H_0]=0$
and the anticommuting relation $\{X_c,H_1\}=0$;  (ii)
The control UDD Hamiltonian  is then applied so that
system evolution is now under a UDD-reduced effective Hamiltonian
thus achieving state protection upto order $N$; (iii)
Depending on the explicit commuting or anticommuting
relations of $X_c$ with $H_0$ and $H_1$, the UDD sequence
efficiently removes a few operators $Y_{i}$ from the initial
generating algebra of $H$ and hence suppresses all couplings
between the state $\vert \chi \rangle$ and all other states.

To protect the general two-qubit
state $\vert \psi \rangle$ in $\cal{P}$
against decoherence using NUDD, it has to be 
locked by nesting three layers of UDD sequences:

\noindent{$\bullet$ \bf Innermost UDD layer:} 
The diagonal populations 
$\text{Tr}[\rho(t) \vert 01\rangle\langle
01 \vert]\approx \vert \alpha \vert^2 $ are locked by
this UDD layer 
with the control operator $X_{0}= I - 2 \vert 01\rangle\langle01 \vert$. 
The reduced effective
Hamiltonian is given by 
$H_{{\rm eff}}^{{\rm UDD-1}} = \sum_{i=1}^{10} D_{1,i} Y_i$,
where $D_{1,i}$ refer to the expansion coefficients of
this first UDD layer. Terms containing basis operators
$Y_{11} \cdots Y_{16}$ are efficiently decoupled.

\noindent{$\bullet$ \bf Second UDD layer:}
The diagonal populations
$\text{Tr}[\rho(t) \vert 10\rangle\langle10
\vert]\approx
\vert \beta \vert^2$ are locked  by this second UDD layer
with the control operator $X_{1}= I -
2 \vert 10\rangle\langle10 \vert$. This UDD sequence is applied to the
reduced effective Hamiltonian 
$H_{{\rm eff}}^{UDD-1}$ 
(defined in the
step above), yielding a further reduced effective
Hamiltonian
$H_{{\rm eff}}^{UDD-2}=\sum_{i=1}^{6} D_{2,i} Y_i$
where $D_{2,i}$ refer to the expansion coefficients of
this second UDD layer. Terms containing basis operators
$Y_{7} \cdots Y_{10}$ are efficiently decoupled.

\noindent{$\bullet$ \bf Outermost UDD layer:}
The off-diagonal coherences
$\text{Tr}[\rho(t) \vert 01\rangle\langle 
10 \vert] \approx \alpha \beta^{*}$ are locked by this
final UDD layer 
with the control operator
$X_{\phi}= I - [\vert 01\rangle+ \vert 10\rangle][\langle01
\vert+\langle
10 \vert]$.
The final reduced effective Hamiltonian after the 
three-layer NUDD contains
five operators:
$H_{{\rm eff}}^{{\rm UDD-3}}=\sum_{i=1}^{5} D_{3,i}Y_i$,
where $D_{3,i}$ are the coefficients due to three UDD layers.

The innermost UDD control $X_0$ pulses are applied at the 
time intervals
$T_{j,k,l}$, the middle layer UDD control $X_1$ pulses are
applied at the time intervals $T_{j,k}$ and the outermost
UDD control $X_{\phi}$ pulses are applied at the time
intervals $T_j$ ($j,k,l=1,2,...N$) given by:
\begin{eqnarray}
T_{j,k,l} &=& T_{j,k} + (T_{j,k+l} - T_{j,k})
\sin^2{\left(\frac{l \pi}{2N+2}\right)} \nonumber \\
T_{j,k} &=& T_{j} + (T_{j+1} - T_{j})
\sin^2{\left(\frac{k \pi}{2N+2}\right)} \nonumber \\
T_{j} &=& T \sin^2{\left(\frac{j \pi}{2N+2}\right)} 
\end{eqnarray}
The total time interval in the $N^{th}$ order sequence is
$(N+1)^3$ with the total number of pulses in one run being
given by $N((N+1)^2+N+2)$ for even
$N$~\cite{mukhtar-pra-10-2}. 
\begin{figure}[h] 
\centering
\includegraphics[angle=0,scale=1]{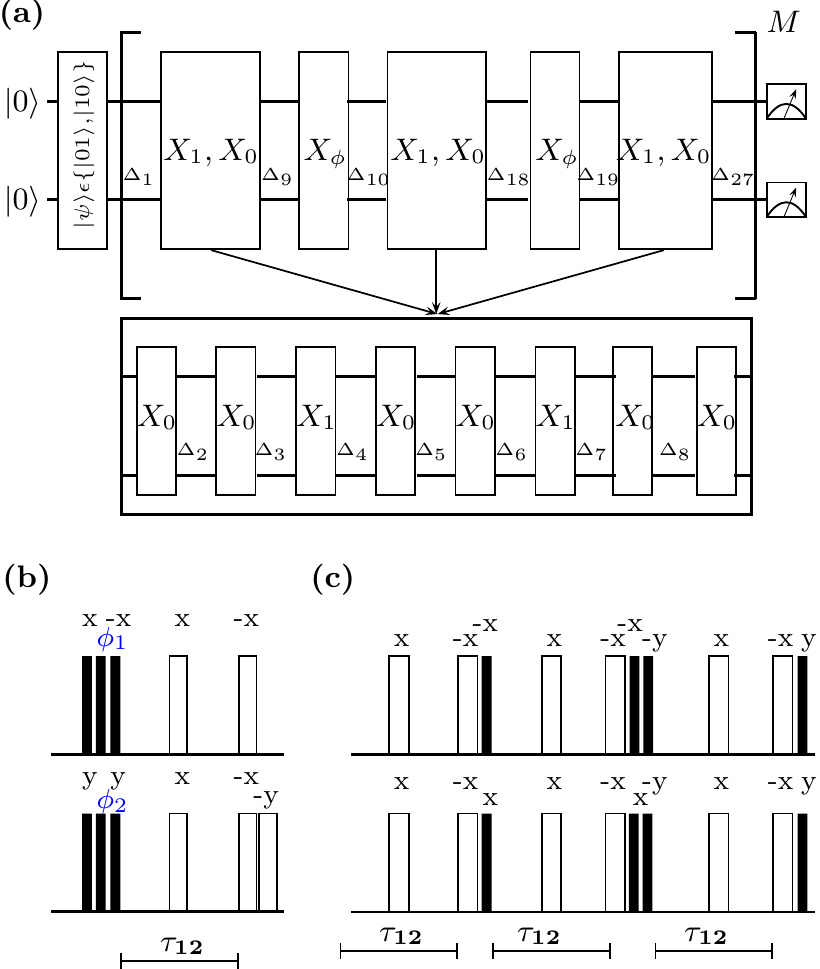} 
\caption{(Color online) (a) Circuit diagram for
the  three-layer NUDD sequence. 
The innermost UDD layer
consists of $X_0$ control pulses, the middle layer
comprises $X_1$ control pulses and the outermost layer
consists of $X_{\phi}$ pulses. The entire NUDD sequence
is repeated $M$ times;
$\Delta_i$ are time intervals.
(b) NMR pulse sequence to implement the control pulses 
for $X_{0}$ and $X_{1}$ UDD sequences. The values of the
rf pulse phases  
$\phi_1$ and $\phi_2$ are set to $x$ and
$y$ for the $X_0$ and to $-x$ and $-y$ for the
$X_1$ UDD sequence, respectively.
(c) NMR pulse sequence to implement the
control pulses for the $X_{\phi}$ UDD sequence.
The filled rectangles
denote $\pi/2$ pulses while the unfilled rectangles
denote $\pi$ pulses, respectively.
The time period $\tau_{12}$ is set to the value $(2
J_{12})^{-1}$, where $J_{12}$ denotes the strength of the
scalar coupling between the two qubits.
}
\label{nudd_ckt}
\end{figure}
\begin{figure}[h]
\includegraphics[angle=0,scale=1.0]{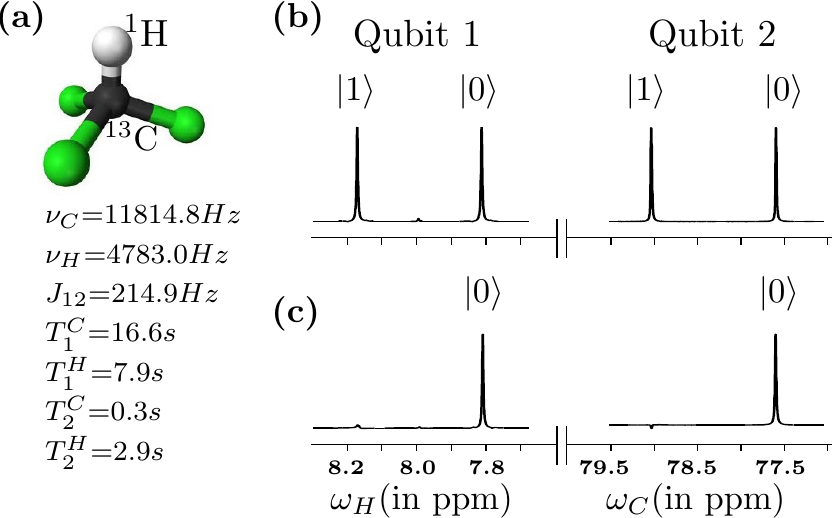}
\caption{(Color online) (a) 
Structure of isotopically enriched 
chloroform-${}^{13}$C molecule, with the
${}^{1}$H spin labeling the first qubit 
and the ${}^{13}$C spin labeling the second qubit.
The system parameters are tabulated alongside
with chemical shifts $\nu_i$ and scalar coupling $J_{12}$ (in
Hz) and NMR spin-lattice
and spin-spin relaxation times $T_{1}$ and $T_{2}$ (in seconds).
(b) 
NMR spectrum obtained after a $\pi/2$
readout pulse on the thermal equilibrium state
and
(c) NMR spectrum of the pseudopure $\vert 0 0 \rangle$
state.
The resonance lines of each qubit in the spectra are labeled by
the corresponding logical states of the other qubit. 
}
\label{molecule}
\end{figure}
\section{Experimental protection of two qubits using NUDD}
\subsection{NMR implementation of NUDD}
\label{expt1}
We now turn to the NUDD implementation for $N=2$ on
a two-qubit NMR system.
The entire NUDD sequence can be written in terms of 
UDD control operators 
$X_0, X_1, X_{\phi}$ (defined in the previous section) and
time evolution $U(\delta_i t)$ under the general Hamiltonian
for time interval fractions $\delta_i$:
\begin{eqnarray}
X_{c}(t)
&=& U(\delta_1t)X_{0}U(\delta_2t)X_{0}U(\delta_3t)X_{1}
U(\delta_4t)X_{0}U(\delta_5t) 
X_{0}
\nonumber \\
&&
U(\delta_6t)X_{1}U(\delta_7t)X_{0}
U(\delta_8t)X_{0}U(\delta_9t) X_{\phi}U(\delta_{10}t) 
X_{0} 
\nonumber \\  
&&
U(\delta_{11}t)
X_{0}U(\delta_{12}t)X_{1}
U(\delta_{13}t)X_{0}U(\delta_{14}t)
X_{0}
U(\delta_{15}t)
\nonumber \\
&&
X_{1}
U(\delta_{16}t)X_{0}
U(\delta_{17}t)X_{0}U(\delta_{18}t)
X_{\phi}
U(\delta_{19}t)
\nonumber \\ 
&&
X_{0}
U(\delta_{20}t)
X_{0}
U(\delta_{21}t)X_{1}U(\delta_{22}t)X_{0}
U(\delta_{23}t)
\nonumber \\ 
&&
X_{0}
U(\delta_{24}t)X_{1}
U(\delta_{25}t)
X_{0}U(\delta_{26}t)X_{0}U(\delta_{27}t) 
\end{eqnarray}
In our implementation, the number of
$X_0, X_1$ and $X_{\phi}$ control pulses used 
in one run of the three-layer NUDD sequence are
18, 6 and 2, respectively.

Using the UDD timing intervals defined above and
applying the condition $\sum \delta_i=1$, 
their values are computed to be
\begin{eqnarray}
\{\delta_i \}
&=&\{\beta, 2\beta, \beta, 2\beta, 4\beta,
2\beta,\beta,2\beta, \beta, 2\beta, 4\beta,
2\beta, 4\beta,8\beta,  \nonumber \\ &&4\beta,
2\beta, 4\beta, 2\beta,\beta, 2\beta, \beta,
2\beta,4\beta, 2\beta,\beta,2\beta,\beta \}
\label{delta}
\end{eqnarray}
where the intervals between the
$X_0, X_1$ and $X_{\phi}$ control pulses 
turn out to be a multiple of $\beta=0.015625 $ .

The NUDD scheme for state protection and the
corresponding NMR pulse sequence is given in
Fig.~\ref{nudd_ckt}.  The unitary gates $X_{0}$,
$X_{1}$, and $X_{\phi}$ drawn in
Fig.~\ref{nudd_ckt}(a) correspond to the UDD
control operators already defined in the previous
section.  The $\Delta_i$ time interval in the
circuit given in Fig.~\ref{nudd_ckt}(a) is defined
by $\Delta_i = \delta_i t$, using the $\delta_i$
given in Eqn.~(\ref{delta}). The pulses on the top
line in Figs.~\ref{nudd_ckt}(b) and (c) are
applied on the first qubit (${}^{1}$H spin in
Fig.~\ref{molecule}), while those at the bottom
are applied on the second qubit (${}^{13}$C spin
in Fig.~\ref{molecule}), respectively.  All the
pulses are spin-selective pulses, with the
$90^{\circ}$ pulse length being $7.6 \mu$s and
$15.6 \mu$s for the proton and carbon rf channels,
respectively.  When applying pulses simultaneously
on both the carbon and proton spins, care was
taken to ensure that the pulses are centered
properly and the delay between two pulses was
measured from the center of the pulse duration
time.  We note here that the NUDD schemes are
experimentally demanding to implement as they
contain long repetitive cycles of rf pulses
applied simultaneously on both qubits and the
timings of the UDD control sequences were matched
carefully with the duty cycle of the rf probe
being used.

We chose the chloroform-${}^{13}$C molecule as the two-qubit
system to implement the NUDD sequence (see
Fig.~\ref{molecule} for details of system parameters and
average NMR relaxation times of both the qubits).
The two-qubit system Hamiltonian in the rotating frame
(which includes the Hamiltonians $H_{S}$ and $H_{12}$ of
Eqn.~(\ref{totham1}))
is given by
\begin{equation}
H_{\rm rot} = -(\nu_H I_{z}^{H} + \nu_C I_{z}^{C}) 
+ 2 \pi J_{12} I_{z}^{H} I_{z}^{C}
\end{equation}
where $\nu_H$ ($\nu_C$) is the chemical shift of the
${}^{1}$H(${}^{13}$C) spin, 
$I_{z}^{H}$($I_z^{C}$) is the 
$z$ component of the spin
angular momentum operator 
for the 
${}^{1}$H(${}^{13}$C) spin, 
and $J_{12}$ is the spin-spin 
scalar coupling
constant. 
The two qubits
were initialized into the pseudopure state $\vert 00 \rangle$
using the spatial averaging technique~\cite{cory-physicad},
with the corresponding
density operator given by
\begin{equation}
\rho_{00} = \frac{1-\epsilon}{4} I
+ \epsilon \vert 00 \rangle \langle 00 \vert
\label{ppure}
\end{equation}
with a thermal polarization $\epsilon \approx
10^{-5}$ and $I$ being a $4 \times 4$
identity operator.  
All experimental density matrices were reconstructed using a
reduced tomographic protocol~\cite{leskowitz-pra-04} and
using the maximum likelihood estimation
technique~\cite{singh-pla-16}.
The fidelity of an experimental
density matrix was computed by 
measuring the
projection between the
theoretically expected and experimentally
measured states using the Uhlmann-Jozsa
fidelity measure~\cite{uhlmann-fidelity,jozsa-fidelity}:
\begin{equation}
F =
\left(Tr \left( \sqrt{
\sqrt{\rho_{\rm theory}}
\rho_{\rm expt} \sqrt{\rho_{\rm theory}}
}
\right)\right)^2
\label{fidelity}
\end{equation}
where $\rho_{\rm theory}$ and $\rho_{\rm expt}$ denote the
theoretical and experimental density matrices
respectively. 
The experimentally created
pseudopure state $\vert 00 \rangle$ was
tomographed with a fidelity of $0.99$.
\begin{figure}[h]
\centering
\includegraphics[angle=0,scale=1.0]{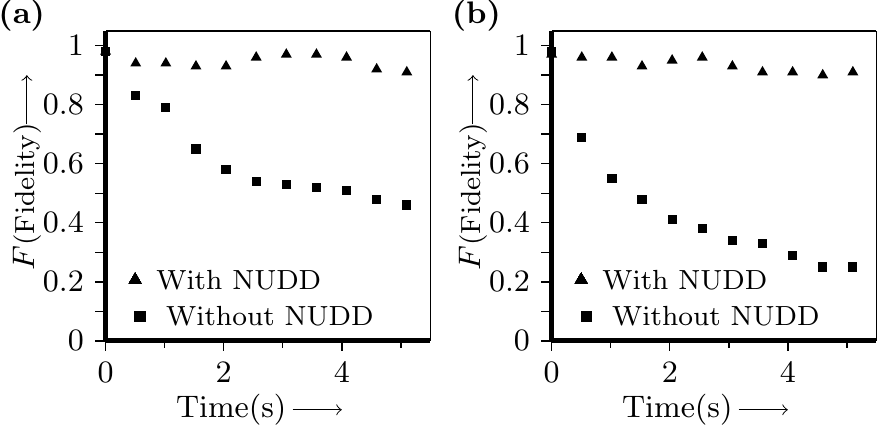}
\caption{Plot of fidelity versus time for (a) the $\vert 01\rangle$
state and (b) the $\vert 10 \rangle)$ state, without any 
protection and after applying NUDD protection. The
fidelity of both the states remains close to 1 for
upto long times, after NUDD
protection.
}
\label{fidelity}
\end{figure}
\subsection{\bf NUDD protection of known states in the
subspace}
\label{known}
We begin evaluating the efficiency of the NUDD scheme by
first applying it to protect four known states in the
two-dimensional subspace ${\cal P}$, namely two separable
and two maximally entangled (Bell) states.

\noindent{\bf Protecting two-qubit separable
states:} We experimentally created the two-qubit
separable states $\vert 01 \rangle$ and $\vert 10
\rangle$ from the initial state $\vert00 \rangle$
by applying a $\pi_x$ on the second qubit and on
the first qubit, respectively.  The states were
prepared with a fidelity of 0.98 and 0.97,
respectively.  One run of the NUDD sequence took
0.12756 s and $t=0.05$s (which included time taken
to implement the control operators).  The entire
NUDD sequence was applied 40 times.  The state
fidelity was computed at different time instants,
without any protection and after applying NUDD
protection.  The state fidelity remains close to
0.9 for long times (upto 5 seconds) when NUDD is
applied, whereas for no protection the $\vert 01
\rangle$ state loses its fidelity (fidelity
approaches 0.5) after 3 s and the $\vert 10
\rangle$ state loses its fidelity after 2 s. A
plot of state fidelities versus time is displayed
in Fig.~\ref{fidelity}, demonstrating the
remarkable efficacy of the NUDD sequence in
protecting separable two-qubit states against
decoherence.
\begin{figure}[h]
\centering
\includegraphics[angle=0,scale=1.0]{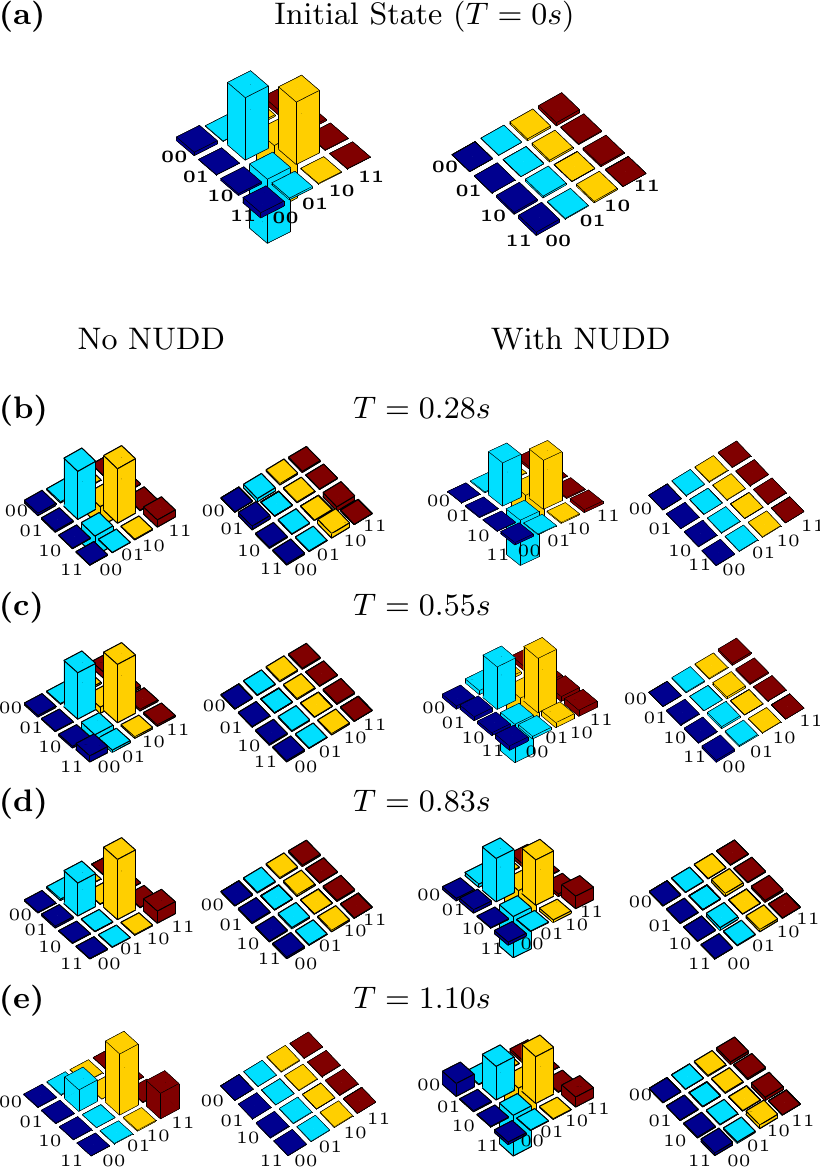}
\caption{(Color online)
Real (left) and imaginary (right) parts of the experimental
tomographs of the (a)
$\frac{1}{\sqrt{2}}(|01\rangle-|10\rangle)$ state, with a
computed fidelity of 0.99.  (b)-(e) depict the state at $T =
0.28, 0.55, 0.83, 1.10$s, with the tomographs on the left
and the right representing the state without any
protection and after
applying NUDD protection, respectively.  The rows
and columns are labeled in the computational basis ordered
from $\vert 00 \rangle$ to $\vert 11 \rangle$.
}
\label{tomosinglet}
\end{figure}
\begin{figure}[h!]
\centering
\includegraphics[angle=0,scale=1.0]{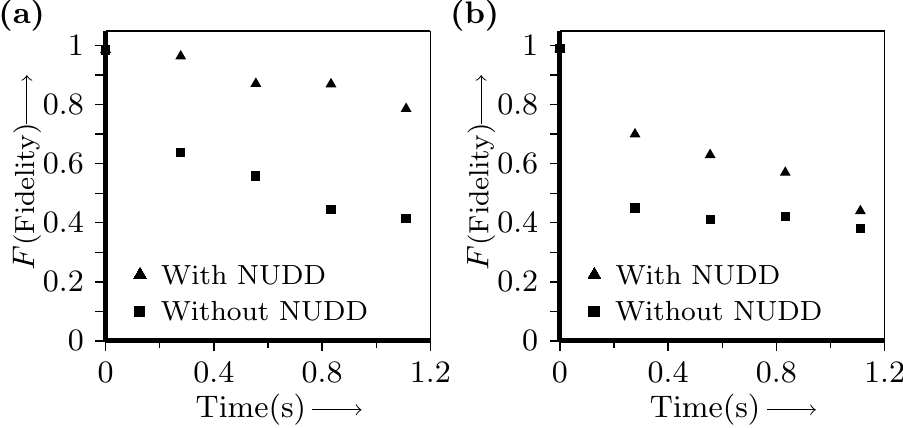}
\caption{Plot of fidelity versus time for (a) the Bell
singlet state and (b) the Bell triplet state, without any 
protection and after applying NUDD protection. 
}
\label{bell-fidelity}
\end{figure}

\begin{figure}[h]
\centering
\includegraphics[angle=0,scale=1.0]{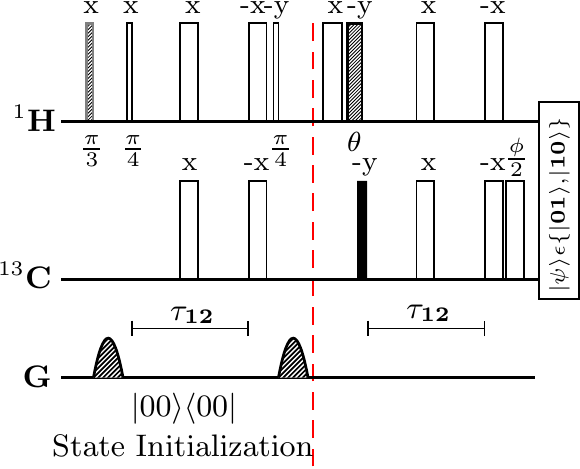}
\caption{(Color online) NMR pulse sequence for the
preparation of random states.  The sequence of pulses before
the vertical dashed red line achieve state initialization
into the $\vert 00 \rangle$ state.  The values of flip
angles $\theta$ and $\phi$ of the rf pulses are randomly
generated.  Filled and unfilled rectangles represent
$\frac{\pi}{2}$ and $\pi$ pulses respectively, while all
other rf pulses are labeled with their respective flip
angles and phases; the interval $\tau_{12}$ is set to $(2
J_{12})^{-1}$ where $J_{12}$ is the scalar
coupling.}
\label{randomckt}
\end{figure}

\noindent{\bf Protecting two-qubit Bell states:}
We next implemented NUDD protection on the 
maximally entangled singlet state
$\frac{1}{\sqrt{2}}(\vert 01\rangle- \vert 10\rangle)$.
We experimentally constructed
the singlet state from the initial
$\vert00 \rangle$ state via the pulse sequence given 
in Fig.~\ref{randomckt} with values of $\theta=-\frac{\pi}{2}$
and $\phi=0$.  
The fidelity of the experimentally constructed singlet
state was computed to be 0.99.
One run of the
NUDD sequence took 0.27756 s
and $t$ was kept at $t=0.2$s.
The
entire NUDD sequence was applied 4 times on the state. 
The singlet state fidelity at different time points was computed 
without any protection and after applying NUDD protection,
and the state tomographs are displayed in
Fig.~\ref{tomosinglet} (tomographs for other states
not shown).
The fidelity of the singlet state
remained close to 0.8 for 1 s when NUDD protection
was applied, whereas 
when no protection is applied the state decoheres
(fidelity approaches 0.5) after 0.55 s.
We also implemented NUDD protection on the 
maximally entangled triplet state
$\frac{1}{\sqrt{2}}(\vert 01\rangle+\vert 10\rangle)$.
We experimentally constructed
the triplet state from the initial
$\vert 00 \rangle$ state via the pulse sequence given 
in Fig.~\ref{randomckt} with values of $\theta=\frac{\pi}{2}$
and $\phi=0$.  
The fidelity of the experimentally constructed triplet
state was computed to be 0.99.
The total NUDD time was kept at 
$t=0.2$s and one run of the
NUDD sequence took 0.27756 s.  The
entire NUDD sequence was repeated 4 times on the state. 
The state fidelity at different time points was computed 
without any protection and after applying NUDD protection.
The fidelity of the triplet state
remained close to 0.8 for 0.28 s when NUDD protection
was applied, whereas 
when no protection is applied the state decoheres
quite rapidly (fidelity approaches 0.5) after 0.28 s.
A plot of state fidelities of both Bell
states versus time
is displayed in Fig.~\ref{bell-fidelity}.
While the NUDD scheme was able to protect the singlet
state quite well (the time for which the state
remains protected is double as compared to its
natural decay time), it is not able to extend the
lifetime of the triplet state to any appreciable extent.
However, what is worth noting here is the fact that the
state fidelity remains close to 0.8 under NUDD protection,
implying that there is no ``leakage'' to other states.
\begin{figure}[b]
\centering
\includegraphics[angle=0,scale=0.9]{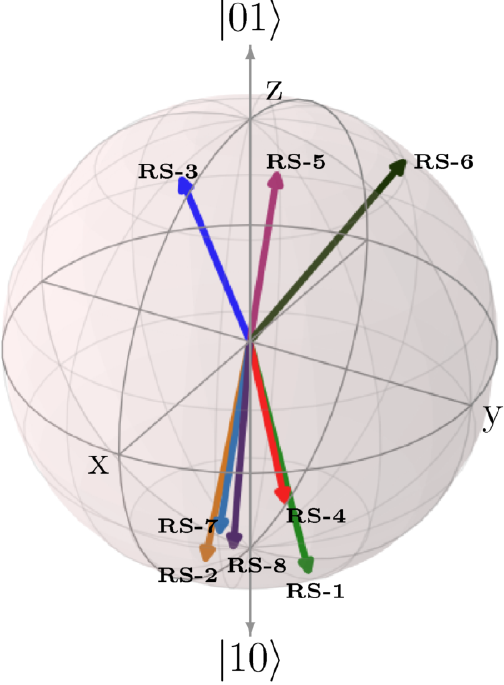}
\caption{(Color online) Geometrical representation of eight randomly 
generated states on a Bloch sphere belonging to the two-qubit
subspace ${\cal P} =\{\vert 01\rangle,\vert 10\rangle\}$.
Each vector makes angles $\theta,\phi$ with the $z$ and
$x$ axes, respectively.
The state labels RS-$i$ ($i=1..8$) are explained in the text.
}
\label{blochfig}
\end{figure}
\subsection{\bf NUDD protection of unknown states in the subspace}
\label{unknown}
We wanted to carry out an unbiased assessment of the
efficacy of the NUDD scheme for state protection. To this
end, we randomly generated several states in the two-dimensional
subspace ${\cal P}$, and applied  the NUDD sequence on each
state.
\begin{table*}[ht]
\caption{\label{table1} 
Results of applying NUDD protection on eight randomly 
generated states in the two-dimensional 
subspace.
Each random state (RS) is tagged with a number for 
convenience, and its corresponding  
($\theta$,$\phi$) angles are given in the column alongside.
The fourth column displays the time at which the state
fidelity approaches 0.5 (an estimate of the natural decay 
time of the state) and the last column displays the time for which
state fidelity remains close to $\approx 0.8$ after 
applying NUDD protection.}
\centering
\begin{ruledtabular}
\begin{tabular}{ccccc}
{\bf State}& {\bf Label} &$(\theta,\phi)$({\bf deg})&
{\bf Decay Time (s)} & {\bf Protected Time (s)} \\
\hline
$ 0.2869|01\rangle + (0.9403 + \iota 0.1828) |10\rangle$&
RS-1&(147,57)&0.5s&1.0s\\
$ 0.1474 |01\rangle - (0.7586 + \iota 0.6346) |10\rangle$&
RS-2&(163,349)&0.5s&1.1s\\
$ 0.9802 |01\rangle + (0.1079 - \iota0.1662) |10\rangle$&
RS-3&(23,345)&1.1s&1.1s\\
$ 0.1356 |01\rangle +  (0.3646 -\iota 0.9212) |10\rangle$&
RS-4&(164,175)&0.6s&1.1s\\
$ 0.9883 |01\rangle +  (0.1048 + \iota0.1109) |10\rangle$&
RS-5&(18,51)&1.1s&1.1s\\
$ 0.9058 |01\rangle + (0.2153 +\iota 0.3648) |10\rangle$&
RS-6&(50,152)&0.6s&0.9s\\
$ 0.0667 |01\rangle + (-0.7693 +\iota 0.6353) |10\rangle$&
RS-7&(172,285)&0.6s&1.1s\\
$ 0.0551 |01\rangle + (0.9861 -\iota 0.1570) |10\rangle$&
RS-8&(174,346)&0.6s&1.1s\\
\end{tabular}
\end{ruledtabular}
\end{table*}
\begin{figure}[h]
\centering
\includegraphics[angle=0,scale=1.0]{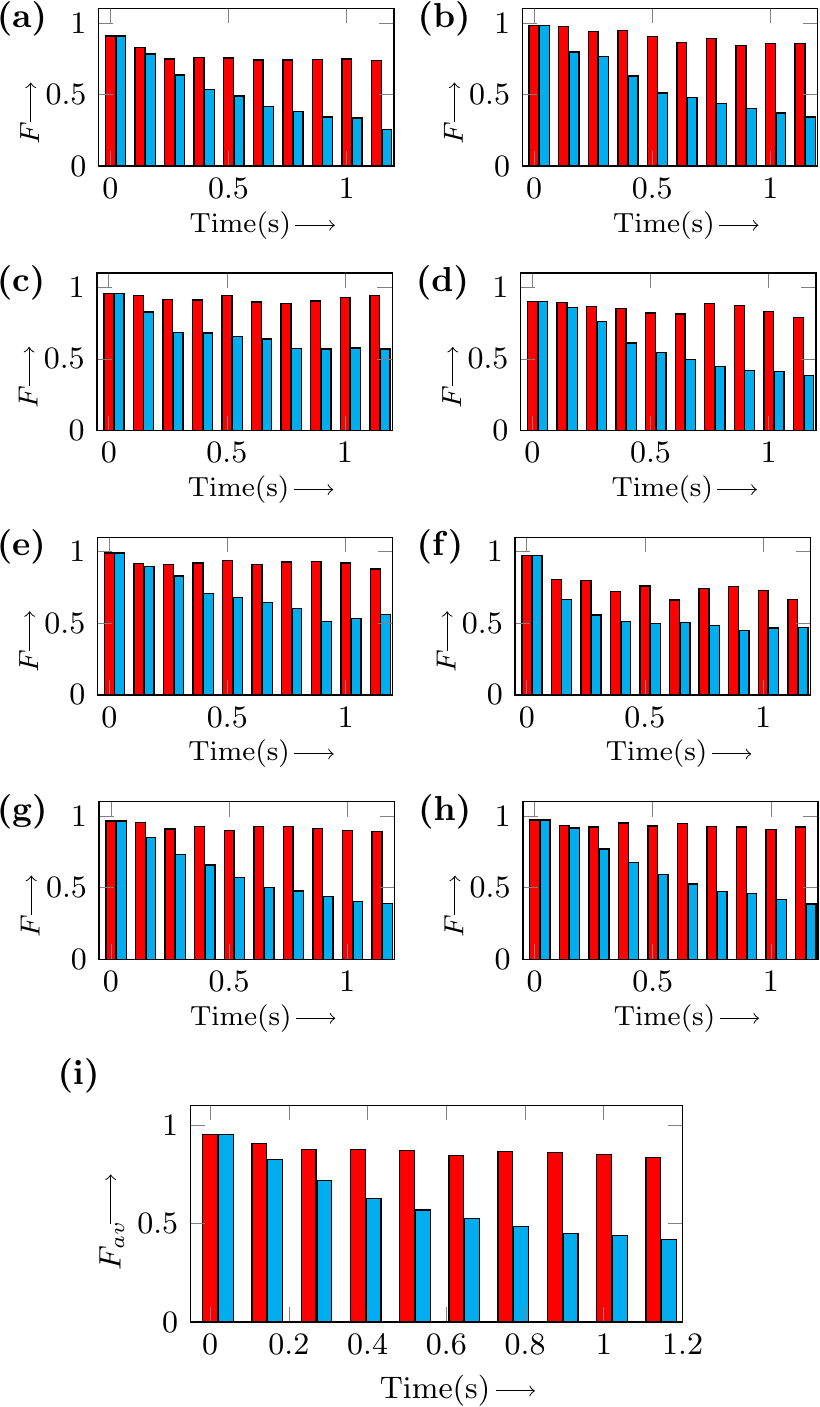}
\caption{(Color online) Bar plots of fidelity versus time of 
eight randomly generated states (labeled RS-$i$, $i=1..8$),
without any protection (blue bars) 
and after applying NUDD protection (red bars):
(a) RS-1, (b) RS-2, (c) RS-3, (d) RS-4, (e) RS-5, 
(f) RS-6, (g) RS-7 and (h) RS-8.
(i) Bar plot showing average fidelity of all eight randomly
generated states, at each time point.
The
state labels are explained in the main text. 
}
\label{random_fid}
\end{figure}
A general state in the two-qubit subspace ${\cal
P} =\{\vert 01\rangle,\vert 10\rangle\}$ can be
written in the form \begin{equation} \vert \psi
\rangle= \cos{\frac{\theta}{2}} \vert
01\rangle+e^{-\iota \phi} \sin{\frac{\theta}{2}}
\vert 10\rangle\ \end{equation} These states were
experimentally created by using random values of
$\theta$ and $\phi$ for the rf pulse flip angles,
as detailed in Fig.~\ref{randomckt}.  The eight
randomly generated two-qubit states are shown in
Fig.~\ref{blochfig}, where the distribution of the
vectors on the Bloch sphere (corresponding to the
two-dimensional subspace ${\cal P}$) shows that
these states are indeed quite random.  The entire
three-layered NUDD sequence was applied 10 times
on each of the eight random states.  The time $t$
for the sequence was kept at $t=0.05$s and one run
of the NUDD sequence took $0.12756$ s.  The plots
of fidelity versus time are shown as bar graphs in
Fig.~\ref{random_fid}, with the blue bars
representing state fidelity without any protection
and the red bars representing state fidelity after
NUDD protection.  The final bar plot in
Fig.~\ref{random_fid}(i) shows the average
fidelity of all the randomly generated states at
each time point.  The results of protecting these
random states via three-layered NUDD are tabulated
in Table~\ref{table1}. Each state has been tagged
by a label RS-$i$ (RS denoting ``Random State''
and $i=1,..8$), with its $\theta,\phi$ values
displayed in the next column.  The fourth column
displays the values of the natural decoherence
time (in seconds) of each state (estimated by
computing the time at which state fidelity
approaches 0.5). The last column in the table
displays the time for which the state remains
protected after applying NUDD (estimated by
computing the time upto which state fidelity
remains close to 0.8).  While the NUDD scheme is
able to protect specific states in the subspace
with varying degrees of success (as evidenced from
the entries in the last column of in
Table~\ref{table1}), on an average as seen from
the bar plot of the average fidelity in
Fig.~\ref{random_fid}(i), the scheme performs
quite well.
\section{Conclusions}
\label{concl}
We experimentally implemented a three-layer nested
UDD sequence on an NMR quantum information processor and explored its
efficiency in protecting arbitrary 
states in a two-dimensional subspace
of two qubits.  The nested UDD layers were applied in a
particular sequence and the full NUDD scheme was
able to achieve second order decoupling of the
system and bath.  The scheme is sufficiently
general as it does not assume prior information
about the explicit form of the system-bath
coupling.  The experiments were highly demanding,
with the control operations being complicated and
involving manipulations of both qubits
simultaneously.  However, our results demonstrate
that such systematic NUDD schemes 
can be experimentally implemented, and are able to
protect multiqubit states in systems that are
arbitrarily coupled to quantum baths.

The beauty of the NUDD schemes lies in the fact
that one is sure the schemes will work to some
extent! Furthermore, one need not know anything
about the state to be protected or the nature of
the quantum channel responsible for its
decoherence. All one needs to know is the subspace
to which the state belongs.  Analogous to an
expert huntswoman who knows her quarry well and
sets her traps accordingly, if the QIP
experimentalist has full knowledge of the state
she wants to protect, she might be served better
by using UDD schemes that are not nested.  However
if the nature of the beast to be captured is
unclear, the QIP experimentalist might do better
by setting a ``generic trap'' such as these NUDD
schemes, knowing that some amount of state
protection will always occur.  Our study points
the way to the realistic protection of fragile
quantum states upto high orders and against
arbitrary noise.
\begin{acknowledgments}
All experiments were performed on a Bruker
Avance-III 600 MHz FT-NMR spectrometer at the NMR
Research Facility at IISER Mohali.  Arvind
acknowledges funding from DST India under grant
number EMR/2014/000297.  KD acknowledges funding
from DST India under grant number EMR/2015/000556.
HS acknowledges financial support from CSIR India. 
\end{acknowledgments}

\begin{thebibliography}{40}%
\makeatletter
\providecommand \@ifxundefined [1]{%
 \@ifx{#1\undefined}
}%
\providecommand \@ifnum [1]{%
 \ifnum #1\expandafter \@firstoftwo
 \else \expandafter \@secondoftwo
 \fi
}%
\providecommand \@ifx [1]{%
 \ifx #1\expandafter \@firstoftwo
 \else \expandafter \@secondoftwo
 \fi
}%
\providecommand \natexlab [1]{#1}%
\providecommand \enquote  [1]{``#1''}%
\providecommand \bibnamefont  [1]{#1}%
\providecommand \bibfnamefont [1]{#1}%
\providecommand \citenamefont [1]{#1}%
\providecommand \href@noop [0]{\@secondoftwo}%
\providecommand \href [0]{\begingroup \@sanitize@url \@href}%
\providecommand \@href[1]{\@@startlink{#1}\@@href}%
\providecommand \@@href[1]{\endgroup#1\@@endlink}%
\providecommand \@sanitize@url [0]{\catcode `\\12\catcode `\$12\catcode
  `\&12\catcode `\#12\catcode `\^12\catcode `\_12\catcode `\%12\relax}%
\providecommand \@@startlink[1]{}%
\providecommand \@@endlink[0]{}%
\providecommand \url  [0]{\begingroup\@sanitize@url \@url }%
\providecommand \@url [1]{\endgroup\@href {#1}{\urlprefix }}%
\providecommand \urlprefix  [0]{URL }%
\providecommand \Eprint [0]{\href }%
\@ifxundefined \urlstyle {%
  \providecommand \doi  [0]{\begingroup \@sanitize@url \@doi}%
  \providecommand \@doi [1]{\endgroup \@@startlink {\doibase
  #1}doi:\discretionary {}{}{}#1\@@endlink }%
}{%
  \providecommand \doi  [0]{doi:\discretionary{}{}{}\begingroup
  \urlstyle{rm}\Url }%
}%
\providecommand \doibase [0]{http://dx.doi.org/}%
\providecommand \Doi [0]{\begingroup \@sanitize@url \@Doi }%
\providecommand \@Doi  [1]{\endgroup\@@startlink{\doibase#1}\@@Doi}%
\providecommand \@@Doi [1]{#1\@@endlink}%
\providecommand \selectlanguage [0]{\@gobble}%
\providecommand \bibinfo  [0]{\@secondoftwo}%
\providecommand \bibfield  [0]{\@secondoftwo}%
\providecommand \translation [1]{[#1]}%
\providecommand \BibitemOpen [0]{}%
\providecommand \bibitemStop [0]{}%
\providecommand \bibitemNoStop [0]{.\EOS\space}%
\providecommand \EOS [0]{\spacefactor3000\relax}%
\providecommand \BibitemShut  [1]{\csname bibitem#1\endcsname}%
\bibitem [{\citenamefont {Viola}(2004)}]{viola-review}%
  \BibitemOpen
  \bibfield  {author} {\bibinfo {author} {\bibfnamefont {L.}~\bibnamefont
  {Viola}},\ }\Doi {10.1080/09500340408231795} {\bibfield  {journal} {\bibinfo
  {journal} {J. Mod. Opt.},\ }\textbf {\bibinfo {volume} {51}},\ \bibinfo
  {pages} {2357} (\bibinfo {year} {2004})}\BibitemShut {NoStop}%
\bibitem [{\citenamefont {Viola}\ \emph {et~al.}(1999)\citenamefont {Viola},
  \citenamefont {Knill},\ and\ \citenamefont {Lloyd}}]{viola-prl-99-2}%
  \BibitemOpen
  \bibfield  {author} {\bibinfo {author} {\bibfnamefont {L.}~\bibnamefont
  {Viola}}, \bibinfo {author} {\bibfnamefont {E.}~\bibnamefont {Knill}}, \ and\
  \bibinfo {author} {\bibfnamefont {S.}~\bibnamefont {Lloyd}},\ }\Doi
  {10.1103/PhysRevLett.82.2417} {\bibfield  {journal} {\bibinfo  {journal}
  {Phys. Rev. Lett.},\ }\textbf {\bibinfo {volume} {82}},\ \bibinfo {pages}
  {2417} (\bibinfo {year} {1999})}\BibitemShut {NoStop}%
\bibitem [{\citenamefont {Carr}\ and\ \citenamefont
  {Purcell}(1954)}]{carr-pr-54}%
  \BibitemOpen
  \bibfield  {author} {\bibinfo {author} {\bibfnamefont {H.}~\bibnamefont
  {Carr}}\ and\ \bibinfo {author} {\bibfnamefont {E.}~\bibnamefont {Purcell}},\
  }\Doi {10.1103/PhysRev.94.630} {\bibfield  {journal} {\bibinfo  {journal}
  {Physical Review},\ }\textbf {\bibinfo {volume} {94}},\ \bibinfo {pages}
  {630} (\bibinfo {year} {1954})}\BibitemShut {NoStop}%
\bibitem [{\citenamefont {Uhrig}(2008)}]{uhrig-njp-08}%
  \BibitemOpen
  \bibfield  {author} {\bibinfo {author} {\bibfnamefont {G.~S.}\ \bibnamefont
  {Uhrig}},\ }\Doi {10.1088/1367-2630/10/8/083024} {\bibfield  {journal}
  {\bibinfo  {journal} {New. J. Phys.},\ }\textbf {\bibinfo {volume} {10}},\
  \bibinfo {pages} {083024} (\bibinfo {year} {2008})}\BibitemShut {NoStop}%
\bibitem [{\citenamefont {Hodgson}\ \emph {et~al.}(2010)\citenamefont
  {Hodgson}, \citenamefont {Viola},\ and\ \citenamefont
  {D'Amico}}]{hodgson-pra-10}%
  \BibitemOpen
  \bibfield  {author} {\bibinfo {author} {\bibfnamefont {T.~E.}\ \bibnamefont
  {Hodgson}}, \bibinfo {author} {\bibfnamefont {L.}~\bibnamefont {Viola}}, \
  and\ \bibinfo {author} {\bibfnamefont {I.}~\bibnamefont {D'Amico}},\ }\Doi
  {10.1103/PhysRevA.81.062321} {\bibfield  {journal} {\bibinfo  {journal}
  {Phys. Rev. A},\ }\textbf {\bibinfo {volume} {81}},\ \bibinfo {pages}
  {062321} (\bibinfo {year} {2010})}\BibitemShut {NoStop}%
\bibitem [{\citenamefont {Schroeder}\ and\ \citenamefont
  {Agarwal}(2011)}]{schroeder-pra-11}%
  \BibitemOpen
  \bibfield  {author} {\bibinfo {author} {\bibfnamefont {C.~A.}\ \bibnamefont
  {Schroeder}}\ and\ \bibinfo {author} {\bibfnamefont {G.~S.}\ \bibnamefont
  {Agarwal}},\ }\Doi {10.1103/PhysRevA.83.012324} {\bibfield  {journal}
  {\bibinfo  {journal} {Phys. Rev. A},\ }\textbf {\bibinfo {volume} {83}},\
  \bibinfo {pages} {012324} (\bibinfo {year} {2011})}\BibitemShut {NoStop}%
\bibitem [{\citenamefont {Yang}\ \emph {et~al.}(2011)\citenamefont {Yang},
  \citenamefont {Wang},\ and\ \citenamefont {Liu}}]{yang-fpc-11}%
  \BibitemOpen
  \bibfield  {author} {\bibinfo {author} {\bibfnamefont {W.}~\bibnamefont
  {Yang}}, \bibinfo {author} {\bibfnamefont {Z.-Y.}\ \bibnamefont {Wang}}, \
  and\ \bibinfo {author} {\bibfnamefont {R.-B.}\ \bibnamefont {Liu}},\ }\Doi
  {10.1007/s11467-010-0113-8} {\bibfield  {journal} {\bibinfo  {journal}
  {Frontiers of Physics in China},\ }\textbf {\bibinfo {volume} {6}},\ \bibinfo
  {pages} {2} (\bibinfo {year} {2011})},\ ISSN \bibinfo {issn}
  {2095-0462}\BibitemShut {NoStop}%
\bibitem [{\citenamefont {Liu}\ \emph {et~al.}(2013)\citenamefont {Liu},
  \citenamefont {Po}, \citenamefont {Du}, \citenamefont {Liu},\ and\
  \citenamefont {Pan}}]{liu-nc-13}%
  \BibitemOpen
  \bibfield  {author} {\bibinfo {author} {\bibfnamefont {G.-Q.}\ \bibnamefont
  {Liu}}, \bibinfo {author} {\bibfnamefont {H.~C.}\ \bibnamefont {Po}},
  \bibinfo {author} {\bibfnamefont {J.}~\bibnamefont {Du}}, \bibinfo {author}
  {\bibfnamefont {R.-B.}\ \bibnamefont {Liu}}, \ and\ \bibinfo {author}
  {\bibfnamefont {X.-Y.}\ \bibnamefont {Pan}},\ }\Doi
  {10.1103/PhysRevLett.82.2417} {\bibfield  {journal} {\bibinfo  {journal} {Nat
  Commun},\ }\textbf {\bibinfo {volume} {4}},\ \bibinfo {pages} {2254}
  (\bibinfo {year} {2013})}\BibitemShut {NoStop}%
\bibitem [{\citenamefont {Dhar}\ \emph {et~al.}(2006)\citenamefont {Dhar},
  \citenamefont {Grover},\ and\ \citenamefont {Roy}}]{dhar-prl-06}%
  \BibitemOpen
  \bibfield  {author} {\bibinfo {author} {\bibfnamefont {D.}~\bibnamefont
  {Dhar}}, \bibinfo {author} {\bibfnamefont {L.~K.}\ \bibnamefont {Grover}}, \
  and\ \bibinfo {author} {\bibfnamefont {S.~M.}\ \bibnamefont {Roy}},\ }\Doi
  {10.1103/PhysRevLett.96.100405} {\bibfield  {journal} {\bibinfo  {journal}
  {Phys. Rev. Lett.},\ }\textbf {\bibinfo {volume} {96}},\ \bibinfo {pages}
  {100405} (\bibinfo {year} {2006})}\BibitemShut {NoStop}%
\bibitem [{\citenamefont {Yang}\ and\ \citenamefont {Liu}(2008)}]{yang-prl-08}%
  \BibitemOpen
  \bibfield  {author} {\bibinfo {author} {\bibfnamefont {W.}~\bibnamefont
  {Yang}}\ and\ \bibinfo {author} {\bibfnamefont {R.-B.}\ \bibnamefont {Liu}},\
  }\Doi {10.1103/PhysRevLett.101.180403} {\bibfield  {journal} {\bibinfo
  {journal} {Phys. Rev. Lett.},\ }\textbf {\bibinfo {volume} {101}},\ \bibinfo
  {pages} {180403} (\bibinfo {year} {2008})}\BibitemShut {NoStop}%
\bibitem [{\citenamefont {Uhrig}(2009)}]{uhrig-prl-09}%
  \BibitemOpen
  \bibfield  {author} {\bibinfo {author} {\bibfnamefont {G.~S.}\ \bibnamefont
  {Uhrig}},\ }\Doi {10.1103/PhysRevLett.102.120502} {\bibfield  {journal}
  {\bibinfo  {journal} {Phys. Rev. Lett.},\ }\textbf {\bibinfo {volume}
  {102}},\ \bibinfo {pages} {120502} (\bibinfo {year} {2009})}\BibitemShut
  {NoStop}%
\bibitem [{\citenamefont {Khodjasteh}\ \emph {et~al.}(2011)\citenamefont
  {Khodjasteh}, \citenamefont {Erdelyi},\ and\ \citenamefont
  {Viola}}]{khodjasteh-pra-11}%
  \BibitemOpen
  \bibfield  {author} {\bibinfo {author} {\bibfnamefont {K.}~\bibnamefont
  {Khodjasteh}}, \bibinfo {author} {\bibfnamefont {T.}~\bibnamefont {Erdelyi}},
  \ and\ \bibinfo {author} {\bibfnamefont {L.}~\bibnamefont {Viola}},\ }\Doi
  {10.1103/PhysRevA.83.020305} {\bibfield  {journal} {\bibinfo  {journal}
  {Phys. Rev. A},\ }\textbf {\bibinfo {volume} {83}},\ \bibinfo {pages}
  {020305} (\bibinfo {year} {2011})}\BibitemShut {NoStop}%
\bibitem [{\citenamefont {Mukhtar}\ \emph
  {et~al.}(2010){\natexlab{a}}\citenamefont {Mukhtar}, \citenamefont {Soh},
  \citenamefont {Saw},\ and\ \citenamefont {Gong}}]{mukhtar-pra-10-1}%
  \BibitemOpen
  \bibfield  {author} {\bibinfo {author} {\bibfnamefont {M.}~\bibnamefont
  {Mukhtar}}, \bibinfo {author} {\bibfnamefont {W.~T.}\ \bibnamefont {Soh}},
  \bibinfo {author} {\bibfnamefont {T.~B.}\ \bibnamefont {Saw}}, \ and\
  \bibinfo {author} {\bibfnamefont {J.}~\bibnamefont {Gong}},\ }\Doi
  {10.1103/PhysRevA.81.012331} {\bibfield  {journal} {\bibinfo  {journal}
  {Phys. Rev. A},\ }\textbf {\bibinfo {volume} {81}},\ \bibinfo {pages}
  {012331} (\bibinfo {year} {2010}{\natexlab{a}})}\BibitemShut {NoStop}%
\bibitem [{\citenamefont {Pan}\ \emph {et~al.}(2011)\citenamefont {Pan},
  \citenamefont {Xi},\ and\ \citenamefont {Gong}}]{pan-jpb-11}%
  \BibitemOpen
  \bibfield  {author} {\bibinfo {author} {\bibfnamefont {Y.}~\bibnamefont
  {Pan}}, \bibinfo {author} {\bibfnamefont {Z.-R.}\ \bibnamefont {Xi}}, \ and\
  \bibinfo {author} {\bibfnamefont {J.}~\bibnamefont {Gong}},\ }\Doi
  {10.1088/0953-4075/44/17/175501} {\bibfield  {journal} {\bibinfo  {journal}
  {J. Phys. B: At. Mol. Opt. Phys},\ }\textbf {\bibinfo {volume} {44}},\
  \bibinfo {pages} {175501} (\bibinfo {year} {2011})}\BibitemShut {NoStop}%
\bibitem [{\citenamefont {Cong}\ \emph {et~al.}(2011)\citenamefont {Cong},
  \citenamefont {Chan},\ and\ \citenamefont {Liu}}]{cong-ijqi-11}%
  \BibitemOpen
  \bibfield  {author} {\bibinfo {author} {\bibfnamefont {S.}~\bibnamefont
  {Cong}}, \bibinfo {author} {\bibfnamefont {L.}~\bibnamefont {Chan}}, \ and\
  \bibinfo {author} {\bibfnamefont {J.}~\bibnamefont {Liu}},\ }\Doi
  {10.1142/S0219749911008428} {\bibfield  {journal} {\bibinfo  {journal}
  {International Journal of Quantum Information},\ }\textbf {\bibinfo {volume}
  {09}},\ \bibinfo {pages} {1599} (\bibinfo {year} {2011})}\BibitemShut
  {NoStop}%
\bibitem [{\citenamefont {\'Alvarez}\ \emph {et~al.}(2012)\citenamefont
  {\'Alvarez}, \citenamefont {Souza},\ and\ \citenamefont
  {Suter}}]{alvarez-pra-12}%
  \BibitemOpen
  \bibfield  {author} {\bibinfo {author} {\bibfnamefont {G.~A.}\ \bibnamefont
  {\'Alvarez}}, \bibinfo {author} {\bibfnamefont {A.~M.}\ \bibnamefont
  {Souza}}, \ and\ \bibinfo {author} {\bibfnamefont {D.}~\bibnamefont
  {Suter}},\ }\Doi {10.1103/PhysRevA.85.052324} {\bibfield  {journal} {\bibinfo
   {journal} {Phys. Rev. A},\ }\textbf {\bibinfo {volume} {85}},\ \bibinfo
  {pages} {052324} (\bibinfo {year} {2012})}\BibitemShut {NoStop}%
\bibitem [{\citenamefont {West}\ and\ \citenamefont
  {Fong}(2012)}]{west-njp-12}%
  \BibitemOpen
  \bibfield  {author} {\bibinfo {author} {\bibfnamefont {J.~R.}\ \bibnamefont
  {West}}\ and\ \bibinfo {author} {\bibfnamefont {B.~H.}\ \bibnamefont
  {Fong}},\ }\Doi {10.1088/1367-2630/14/8/083002} {\bibfield  {journal}
  {\bibinfo  {journal} {New. J. Phys.},\ }\textbf {\bibinfo {volume} {14}},\
  \bibinfo {pages} {083002} (\bibinfo {year} {2012})}\BibitemShut {NoStop}%
\bibitem [{\citenamefont {Ahmed}\ \emph {et~al.}(2013)\citenamefont {Ahmed},
  \citenamefont {\'Alvarez},\ and\ \citenamefont {Suter}}]{ahmed-pra-13}%
  \BibitemOpen
  \bibfield  {author} {\bibinfo {author} {\bibfnamefont {M.~A.~A.}\
  \bibnamefont {Ahmed}}, \bibinfo {author} {\bibfnamefont {G.~A.}\ \bibnamefont
  {\'Alvarez}}, \ and\ \bibinfo {author} {\bibfnamefont {D.}~\bibnamefont
  {Suter}},\ }\Doi {10.1103/PhysRevA.87.042309} {\bibfield  {journal} {\bibinfo
   {journal} {Phys. Rev. A},\ }\textbf {\bibinfo {volume} {87}},\ \bibinfo
  {pages} {042309} (\bibinfo {year} {2013})}\BibitemShut {NoStop}%
\bibitem [{\citenamefont {Agarwal}(2010)}]{agarwal-scripta}%
  \BibitemOpen
  \bibfield  {author} {\bibinfo {author} {\bibfnamefont {G.~S.}\ \bibnamefont
  {Agarwal}},\ }\Doi {10.1088/0031-8949/82/03/038103} {\bibfield  {journal}
  {\bibinfo  {journal} {Physica Scripta},\ }\textbf {\bibinfo {volume} {82}},\
  \bibinfo {pages} {038103} (\bibinfo {year} {2010})}\BibitemShut {NoStop}%
\bibitem [{\citenamefont {Song}\ \emph {et~al.}(2013)\citenamefont {Song},
  \citenamefont {Pan},\ and\ \citenamefont {Zairong}}]{song-ijqi-13}%
  \BibitemOpen
  \bibfield  {author} {\bibinfo {author} {\bibfnamefont {H.}~\bibnamefont
  {Song}}, \bibinfo {author} {\bibfnamefont {Y.}~\bibnamefont {Pan}}, \ and\
  \bibinfo {author} {\bibfnamefont {X.}~\bibnamefont {Zairong}},\ }\Doi
  {10.1142/S0219749913500123} {\bibfield  {journal} {\bibinfo  {journal}
  {International Journal of Quantum Information},\ }\textbf {\bibinfo {volume}
  {11}},\ \bibinfo {pages} {1350012} (\bibinfo {year} {2013})}\BibitemShut
  {NoStop}%
\bibitem [{\citenamefont {Franco}\ \emph {et~al.}(2014)\citenamefont {Franco},
  \citenamefont {D'Arrigo}, \citenamefont {Falci}, \citenamefont {Compagno},\
  and\ \citenamefont {Paladino}}]{lofranco-prb-2014}%
  \BibitemOpen
  \bibfield  {author} {\bibinfo {author} {\bibfnamefont {R.~L.}\ \bibnamefont
  {Franco}}, \bibinfo {author} {\bibfnamefont {A.}~\bibnamefont {D'Arrigo}},
  \bibinfo {author} {\bibfnamefont {G.}~\bibnamefont {Falci}}, \bibinfo
  {author} {\bibfnamefont {G.}~\bibnamefont {Compagno}}, \ and\ \bibinfo
  {author} {\bibfnamefont {E.}~\bibnamefont {Paladino}},\ }\Doi
  {10.1103/PhysRevB.90.054304} {\bibfield  {journal} {\bibinfo  {journal}
  {Phys. Rev. B},\ }\textbf {\bibinfo {volume} {90}},\ \bibinfo {pages}
  {054304} (\bibinfo {year} {2014})}\BibitemShut {NoStop}%
\bibitem [{\citenamefont {Biercuk}\ \emph {et~al.}(2009)\citenamefont
  {Biercuk}, \citenamefont {Uys}, \citenamefont {Vandevender}, \citenamefont
  {Shiga}, \citenamefont {Itano},\ and\ \citenamefont
  {Bollinger}}]{biercuk-pra-09}%
  \BibitemOpen
  \bibfield  {author} {\bibinfo {author} {\bibfnamefont {M.}~\bibnamefont
  {Biercuk}}, \bibinfo {author} {\bibfnamefont {H.}~\bibnamefont {Uys}},
  \bibinfo {author} {\bibfnamefont {A.}~\bibnamefont {Vandevender}}, \bibinfo
  {author} {\bibfnamefont {N.}~\bibnamefont {Shiga}}, \bibinfo {author}
  {\bibfnamefont {W.}~\bibnamefont {Itano}}, \ and\ \bibinfo {author}
  {\bibfnamefont {J.}~\bibnamefont {Bollinger}},\ }\Doi
  {10.1103/PhysRevA.79.062324} {\bibfield  {journal} {\bibinfo  {journal}
  {Phys. Rev. A},\ }\textbf {\bibinfo {volume} {79}},\ \bibinfo {pages}
  {062324} (\bibinfo {year} {2009})}\BibitemShut {NoStop}%
\bibitem [{\citenamefont {Szwer}\ \emph {et~al.}(2011)\citenamefont {Szwer},
  \citenamefont {Webster}, \citenamefont {Steane},\ and\ \citenamefont
  {Lucas}}]{szwer-jpb-11}%
  \BibitemOpen
  \bibfield  {author} {\bibinfo {author} {\bibfnamefont {D.~J.}\ \bibnamefont
  {Szwer}}, \bibinfo {author} {\bibfnamefont {S.~C.}\ \bibnamefont {Webster}},
  \bibinfo {author} {\bibfnamefont {A.~M.}\ \bibnamefont {Steane}}, \ and\
  \bibinfo {author} {\bibfnamefont {D.~M.}\ \bibnamefont {Lucas}},\ }\Doi
  {10.1088/0953-4075/44/2/025501} {\bibfield  {journal} {\bibinfo  {journal}
  {J. Phys. B: At. Mol. Opt. Phys},\ }\textbf {\bibinfo {volume} {44}},\
  \bibinfo {pages} {025501} (\bibinfo {year} {2011})}\BibitemShut {NoStop}%
\bibitem [{\citenamefont {Du}\ \emph {et~al.}(2009)\citenamefont {Du},
  \citenamefont {Rong}, \citenamefont {Zhao}, \citenamefont {Wang},
  \citenamefont {Yang},\ and\ \citenamefont {Liu}}]{du-nature-09}%
  \BibitemOpen
  \bibfield  {author} {\bibinfo {author} {\bibfnamefont {J.}~\bibnamefont
  {Du}}, \bibinfo {author} {\bibfnamefont {X.}~\bibnamefont {Rong}}, \bibinfo
  {author} {\bibfnamefont {N.}~\bibnamefont {Zhao}}, \bibinfo {author}
  {\bibfnamefont {Y.}~\bibnamefont {Wang}}, \bibinfo {author} {\bibfnamefont
  {J.}~\bibnamefont {Yang}}, \ and\ \bibinfo {author} {\bibfnamefont
  {R.}~\bibnamefont {Liu}},\ }\Doi {10.1038/nature08470} {\bibfield  {journal}
  {\bibinfo  {journal} {Nature},\ }\textbf {\bibinfo {volume} {461}},\ \bibinfo
  {pages} {1265} (\bibinfo {year} {2009})}\BibitemShut {NoStop}%
\bibitem [{\citenamefont {\'Alvarez}\ \emph {et~al.}(2010)\citenamefont
  {\'Alvarez}, \citenamefont {Ajoy}, \citenamefont {Peng},\ and\ \citenamefont
  {Suter}}]{alvarez-pra-10}%
  \BibitemOpen
  \bibfield  {author} {\bibinfo {author} {\bibfnamefont {G.~A.}\ \bibnamefont
  {\'Alvarez}}, \bibinfo {author} {\bibfnamefont {A.}~\bibnamefont {Ajoy}},
  \bibinfo {author} {\bibfnamefont {X.}~\bibnamefont {Peng}}, \ and\ \bibinfo
  {author} {\bibfnamefont {D.}~\bibnamefont {Suter}},\ }\Doi
  {10.1103/PhysRevA.82.042306} {\bibfield  {journal} {\bibinfo  {journal}
  {Phys. Rev. A},\ }\textbf {\bibinfo {volume} {82}},\ \bibinfo {pages}
  {042306} (\bibinfo {year} {2010})}\BibitemShut {NoStop}%
\bibitem [{\citenamefont {Ajoy}\ \emph {et~al.}(2011)\citenamefont {Ajoy},
  \citenamefont {\'Alvarez},\ and\ \citenamefont {Suter}}]{ajoy-pra-11}%
  \BibitemOpen
  \bibfield  {author} {\bibinfo {author} {\bibfnamefont {A.}~\bibnamefont
  {Ajoy}}, \bibinfo {author} {\bibfnamefont {G.~A.}\ \bibnamefont {\'Alvarez}},
  \ and\ \bibinfo {author} {\bibfnamefont {D.}~\bibnamefont {Suter}},\ }\Doi
  {10.1103/PhysRevA.83.032303} {\bibfield  {journal} {\bibinfo  {journal}
  {Phys. Rev. A},\ }\textbf {\bibinfo {volume} {83}},\ \bibinfo {pages}
  {032303} (\bibinfo {year} {2011})}\BibitemShut {NoStop}%
\bibitem [{\citenamefont {Roy}\ \emph {et~al.}(2011)\citenamefont {Roy},
  \citenamefont {Mahesh},\ and\ \citenamefont {Agarwal}}]{roy-pra-2011}%
  \BibitemOpen
  \bibfield  {author} {\bibinfo {author} {\bibfnamefont {S.~S.}\ \bibnamefont
  {Roy}}, \bibinfo {author} {\bibfnamefont {T.~S.}\ \bibnamefont {Mahesh}}, \
  and\ \bibinfo {author} {\bibfnamefont {G.~S.}\ \bibnamefont {Agarwal}},\
  }\Doi {10.1103/PhysRevA.83.062326} {\bibfield  {journal} {\bibinfo  {journal}
  {Phys. Rev. A},\ }\textbf {\bibinfo {volume} {83}},\ \bibinfo {pages}
  {062326} (\bibinfo {year} {2011})}\BibitemShut {NoStop}%
\bibitem [{\citenamefont {Singh}\ \emph {et~al.}(2014)\citenamefont {Singh},
  \citenamefont {Arvind},\ and\ \citenamefont {Dorai}}]{singh-pra-14}%
  \BibitemOpen
  \bibfield  {author} {\bibinfo {author} {\bibfnamefont {H.}~\bibnamefont
  {Singh}}, \bibinfo {author} {\bibnamefont {Arvind}}, \ and\ \bibinfo {author}
  {\bibfnamefont {K.}~\bibnamefont {Dorai}},\ }\Doi
  {10.1103/PhysRevA.90.052329} {\bibfield  {journal} {\bibinfo  {journal}
  {Phys. Rev. A},\ }\textbf {\bibinfo {volume} {90}},\ \bibinfo {pages}
  {052329} (\bibinfo {year} {2014})}\BibitemShut {NoStop}%
\bibitem [{\citenamefont {Zhang}\ \emph {et~al.}(2014)\citenamefont {Zhang},
  \citenamefont {Souza}, \citenamefont {Brandao},\ and\ \citenamefont
  {Suter}}]{zhang-prl-14}%
  \BibitemOpen
  \bibfield  {author} {\bibinfo {author} {\bibfnamefont {J.}~\bibnamefont
  {Zhang}}, \bibinfo {author} {\bibfnamefont {A.~M.}\ \bibnamefont {Souza}},
  \bibinfo {author} {\bibfnamefont {F.~D.}\ \bibnamefont {Brandao}}, \ and\
  \bibinfo {author} {\bibfnamefont {D.}~\bibnamefont {Suter}},\ }\Doi
  {10.1103/PhysRevLett.112.050502} {\bibfield  {journal} {\bibinfo  {journal}
  {Phys. Rev. Lett.},\ }\textbf {\bibinfo {volume} {112}},\ \bibinfo {pages}
  {050502} (\bibinfo {year} {2014})}\BibitemShut {NoStop}%
\bibitem [{\citenamefont {Jenista}\ \emph {et~al.}(2009)\citenamefont
  {Jenista}, \citenamefont {Stokes}, \citenamefont {Branca},\ and\
  \citenamefont {Warren}}]{jenista-jcp-09}%
  \BibitemOpen
  \bibfield  {author} {\bibinfo {author} {\bibfnamefont {E.~R.}\ \bibnamefont
  {Jenista}}, \bibinfo {author} {\bibfnamefont {A.~M.}\ \bibnamefont {Stokes}},
  \bibinfo {author} {\bibfnamefont {R.~T.}\ \bibnamefont {Branca}}, \ and\
  \bibinfo {author} {\bibfnamefont {W.~S.}\ \bibnamefont {Warren}},\ }\Doi
  {10.1063/1.3263196} {\bibfield  {journal} {\bibinfo  {journal} {J. Chem.
  Phys.},\ }\textbf {\bibinfo {volume} {131}},\ \bibinfo {pages} {204510}
  (\bibinfo {year} {2009})}\BibitemShut {NoStop}%
\bibitem [{\citenamefont {Álvarez}\ \emph {et~al.}(2014)\citenamefont
  {Álvarez}, \citenamefont {Shemesh},\ and\ \citenamefont
  {Frydman}}]{alvarez-jcp-14}%
  \BibitemOpen
  \bibfield  {author} {\bibinfo {author} {\bibfnamefont {G.~A.}\ \bibnamefont
  {Álvarez}}, \bibinfo {author} {\bibfnamefont {N.}~\bibnamefont {Shemesh}}, \
  and\ \bibinfo {author} {\bibfnamefont {L.}~\bibnamefont {Frydman}},\ }\Doi
  {http://dx.doi.org/10.1063/1.4865335} {\bibfield  {journal} {\bibinfo
  {journal} {J. Chem. Phys.},\ }\textbf {\bibinfo {volume} {140}} (\bibinfo
  {year} {2014})},\ \doi {http://dx.doi.org/10.1063/1.4865335}\BibitemShut
  {NoStop}%
\bibitem [{\citenamefont {Kuo}\ \emph {et~al.}(2012)\citenamefont {Kuo},
  \citenamefont {Quiroz}, \citenamefont {Paz-Silva},\ and\ \citenamefont
  {Lidar}}]{kuo-jmp-12}%
  \BibitemOpen
  \bibfield  {author} {\bibinfo {author} {\bibfnamefont {W.}~\bibnamefont
  {Kuo}}, \bibinfo {author} {\bibfnamefont {G.}~\bibnamefont {Quiroz}},
  \bibinfo {author} {\bibfnamefont {G.}~\bibnamefont {Paz-Silva}}, \ and\
  \bibinfo {author} {\bibfnamefont {D.}~\bibnamefont {Lidar}},\ }\Doi
  {10.1063/1.4769382} {\bibfield  {journal} {\bibinfo  {journal} {J. Math.
  Phys.},\ }\textbf {\bibinfo {volume} {53}},\ \bibinfo {pages} {122207}
  (\bibinfo {year} {2012})}\BibitemShut {NoStop}%
\bibitem [{\citenamefont {Wang}\ and\ \citenamefont {Liu}(2011)}]{wang-pra-11}%
  \BibitemOpen
  \bibfield  {author} {\bibinfo {author} {\bibfnamefont {Z.-Y.}\ \bibnamefont
  {Wang}}\ and\ \bibinfo {author} {\bibfnamefont {R.-B.}\ \bibnamefont {Liu}},\
  }\Doi {10.1103/PhysRevA.83.022306} {\bibfield  {journal} {\bibinfo  {journal}
  {Phys. Rev. A},\ }\textbf {\bibinfo {volume} {83}},\ \bibinfo {pages}
  {022306} (\bibinfo {year} {2011})}\BibitemShut {NoStop}%
\bibitem [{\citenamefont {Jiang}\ and\ \citenamefont
  {Imambekov}(2011)}]{jiang-pra-11}%
  \BibitemOpen
  \bibfield  {author} {\bibinfo {author} {\bibfnamefont {L.}~\bibnamefont
  {Jiang}}\ and\ \bibinfo {author} {\bibfnamefont {A.}~\bibnamefont
  {Imambekov}},\ }\Doi {10.1103/PhysRevA.84.060302} {\bibfield  {journal}
  {\bibinfo  {journal} {Phys. Rev. A},\ }\textbf {\bibinfo {volume} {84}},\
  \bibinfo {pages} {060302} (\bibinfo {year} {2011})}\BibitemShut {NoStop}%
\bibitem [{\citenamefont {Mukhtar}\ \emph
  {et~al.}(2010){\natexlab{b}}\citenamefont {Mukhtar}, \citenamefont {Soh},
  \citenamefont {Saw},\ and\ \citenamefont {Gong}}]{mukhtar-pra-10-2}%
  \BibitemOpen
  \bibfield  {author} {\bibinfo {author} {\bibfnamefont {M.}~\bibnamefont
  {Mukhtar}}, \bibinfo {author} {\bibfnamefont {W.~T.}\ \bibnamefont {Soh}},
  \bibinfo {author} {\bibfnamefont {T.~B.}\ \bibnamefont {Saw}}, \ and\
  \bibinfo {author} {\bibfnamefont {J.}~\bibnamefont {Gong}},\ }\Doi
  {10.1103/PhysRevA.82.052338} {\bibfield  {journal} {\bibinfo  {journal}
  {Phys. Rev. A},\ }\textbf {\bibinfo {volume} {82}},\ \bibinfo {pages}
  {052338} (\bibinfo {year} {2010}{\natexlab{b}})}\BibitemShut {NoStop}%
\bibitem [{\citenamefont {Cory}\ \emph {et~al.}(1998)\citenamefont {Cory},
  \citenamefont {Price},\ and\ \citenamefont {Havel}}]{cory-physicad}%
  \BibitemOpen
  \bibfield  {author} {\bibinfo {author} {\bibfnamefont {D.}~\bibnamefont
  {Cory}}, \bibinfo {author} {\bibfnamefont {M.}~\bibnamefont {Price}}, \ and\
  \bibinfo {author} {\bibfnamefont {T.}~\bibnamefont {Havel}},\ }\Doi
  {http://dx.doi.org/10.1016/S0167-2789(98)00046-3} {\bibfield  {journal}
  {\bibinfo  {journal} {Physica D},\ }\textbf {\bibinfo {volume} {120}},\
  \bibinfo {pages} {82} (\bibinfo {year} {1998})}\BibitemShut {NoStop}%
\bibitem [{\citenamefont {Leskowitz}\ and\ \citenamefont
  {Mueller}(2004)}]{leskowitz-pra-04}%
  \BibitemOpen
  \bibfield  {author} {\bibinfo {author} {\bibfnamefont {G.~M.}\ \bibnamefont
  {Leskowitz}}\ and\ \bibinfo {author} {\bibfnamefont {L.~J.}\ \bibnamefont
  {Mueller}},\ }\Doi {10.1103/PhysRevA.69.052302} {\bibfield  {journal}
  {\bibinfo  {journal} {Phys. Rev. A},\ }\textbf {\bibinfo {volume} {69}},\
  \bibinfo {pages} {052302} (\bibinfo {year} {2004})}\BibitemShut {NoStop}%
\bibitem [{\citenamefont {Singh}\ \emph {et~al.}(2016)\citenamefont {Singh},
  \citenamefont {Arvind},\ and\ \citenamefont {Dorai}}]{singh-pla-16}%
  \BibitemOpen
  \bibfield  {author} {\bibinfo {author} {\bibfnamefont {H.}~\bibnamefont
  {Singh}}, \bibinfo {author} {\bibnamefont {Arvind}}, \ and\ \bibinfo {author}
  {\bibfnamefont {K.}~\bibnamefont {Dorai}},\ }\Doi
  {http://dx.doi.org/10.1016/j.physleta.2016.07.046} {\bibfield  {journal}
  {\bibinfo  {journal} {Phys. Lett. A},\ }\textbf {\bibinfo {volume} {380}},\
  \bibinfo {pages} {3051} (\bibinfo {year} {2016})}\BibitemShut {NoStop}%
\bibitem [{\citenamefont {Uhlmann}(1976)}]{uhlmann-fidelity}%
  \BibitemOpen
  \bibfield  {author} {\bibinfo {author} {\bibfnamefont {A.}~\bibnamefont
  {Uhlmann}},\ }\Doi {http://dx.doi.org/10.1016/0034-4877(76)90060-4}
  {\bibfield  {journal} {\bibinfo  {journal} {Rep. Math. Phys.},\ }\textbf
  {\bibinfo {volume} {9}},\ \bibinfo {pages} {273} (\bibinfo {year}
  {1976})}\BibitemShut {NoStop}%
\bibitem [{\citenamefont {Jozsa}(1994)}]{jozsa-fidelity}%
  \BibitemOpen
  \bibfield  {author} {\bibinfo {author} {\bibfnamefont {R.}~\bibnamefont
  {Jozsa}},\ }\Doi {10.1080/09500349414552171} {\bibfield  {journal} {\bibinfo
  {journal} {J. Mod. Opt.},\ }\textbf {\bibinfo {volume} {41}},\ \bibinfo
  {pages} {2315} (\bibinfo {year} {1994})}\BibitemShut {NoStop}%
\end{thebibliography}
\end{document}